# Pressure-induced structural disordering and anomalous pressure-volume behaviour in high-entropy zirconates

Yogendar Singh[1], Xinghua Su[2,3], Vivek Kumar[1], H. K. Poswal[4], K. K. Pandey[4,5*], Pawan Kumar Kulriya[1*]

[1]School of Physical Sciences, Jawaharlal Nehru University, New Delhi 110067, India

[2]School of Materials Science and Engineering, Chang'an University, Xi'an 710061, China

[3]State Key Laboratory of Environment-Friendly Energy Materials, Southwest University of Science and Technology, Mianyang 621010, China

[4]High-Pressure and Synchrotron Radiation Physics Division, Bhabha Atomic Research Centre, Mumbai 400085, India

[5]Homi Bhabha National Institute, Mumbai 400094, India

**Abstract**

The ambient-temperature high-pressure behaviour of $(La_{0.2}Nd_{0.2}Sm_{0.2}Gd_{0.2}Yb_{0.2})_2Zr_2O_7$ zirconate (HEZ) nanopowders with three different average particle sizes (~25nm, ~45 nm and ~ 68nm) were studied using synchrotron X-ray diffraction (SR-XRD) measurements up to ~30 GPa. Smaller particle-size HEZ nanopowder (~25 nm), synthesized at the lower sintering temperature, exhibits pure defect-fluorite (DF) phase, whereas larger particle-size HEZ nanopowders (~45nm and ~68nm), synthesized at the higher sintering temperature, exhibit mixture of DF and pyrochlore phase (PY). The phase fraction of the PY phase increases with sintering temperature and hence with the particle size. All the HEZ nanopowders exhibit stability of initial structures (DF and PY) up to ~ 30 GPa, though phase fraction of PY phase in larger particle-size HEZ nanopowders successively reduces with pressure which is concomitant with significant variation in $o_{x48f}$ fractional coordinate in PY phase. Both the phases in all the studied samples exhibit anomalous pressure-volume (P-V) behaviour between ~7 to 15 GPa. The anomaly decreases with increasing particle size of HEZ nanopowders. The variation of bond lengths and polyhedron volume with pressure suggests that the anomalous P-V behaviour and structural changes at high pressures are primarily due to the distortion of the polyhedrons in DF and PY structures in HEZ nanopowders.

*Authors to whom correspondence should be addressed.

1. **K. K. Pandey -** Homi Bhabha National Institute, Mumbai 400094, India, electronic mail: kkpandey@barc.gov.in

2. **Pawan Kumar Kulriya-** School of Physical Sciences, Jawaharlal Nehru University, New Delhi 110067, India, electronic mail: pkkulriya@mail.jnu.ac.in

# 1. Introduction

Recently, high-entropy materials, particularly high entropy oxides (HEOs) and alloys (HEAs) have attracted significant scientific interest due to their highly customizable characteristics and various technological applications[1–3]. The concept of HEOs was inherited from the field of high entropy alloys, and after synthesizing $(Mg_{0.2}Zn_{0.2}Cu_{0.2}Co_{0.2}Ni_{0.2})O$ with a rock salt crystal structure[4] in 2015. Various compositions of HEOs, including perovskites[5], pyrochlore oxides[6], carbides[7], and orides[8], have been developed in the past few years. These high-entropy ceramics and their derivates typically have certain excellent properties such as superior mechanical strength, flexible crystalline structure, extremely low thermal conductivity, and outstanding electrochemical properties[9,10]. The mixing entropy associated with the homogenous mixing of five or more cations on a single crystallographic site is predicted to enhance the thermodynamic stability in single-phase HEOs.

In contrast to traditional materials, where one or a few constituents dominate, these materials have a higher entropy since all the elements are equally dispersed throughout their structure. Among these, pyrochlore oxides are promising candidates for various technological applications, especially for inert matrix materials and host matrix for nuclear waste immobilization[11]. The high-entropy design idea involves various compositions by substituting cations at A and B sites, making it feasible to extend the pyrochlore solid solution to the high-entropy pyrochlore phase. This could have more functional tunability or emergent characteristics than conventional pyrochlores that have one or two cations at main crystallographic sites.

It is generally known that high pressure[12] and ion radiation[13] can lead to amorphization or structural modifications in many conventional compounds, especially those with polyhedral network crystal structures[11]. Point defects are produced when these pyrochlore compositions are subjected to high pressures, high temperatures, and ion irradiation. The most common type of cation defect is an anti-site defect, in which the sites of A and B cations switch, while the predominant type of anion defect is a Frenkel pair[14,15]. Anion Frenkel-pair formation energy is lowered at high temperatures by creating cation anti-site defects. Anions can move readily due to cation substitution, which also makes all locations on the lattice appear electrically similar[16,17]. At high pressures, anion disorder often occurs at several GPa lower than the cation disorder. In addition to the relative ease of oxygen hopping at high pressures, this disordering is caused by the distortion of polyhedrons to accommodate pressure-induced stress[18,19]. Xu *et al.* studied the response of mechanical properties and structure of high entropy $(La_{0.2}Ce_{0.2}Nd_{0.2}Sm_{0.2}Gd_{0.2})_2Zr_2O_7$ pyrochlore at the extreme conditions of ion irradiation[20]. They found that high entropy pyrochlore exhibits fewer structural modifications than conventional $Nd_2Zr_2O_7$ pyrochlore because of severe lattice distortion and compositional complexity that cause higher migration barriers of Frankel defects aggregation[20]. Furthermore, irradiated high entropy pyrochlores showed improved mechanical properties, resulting from residual stress from irradiation on the surface of the ceramics[20]. There are very limited high-pressure behavior studies of high entropy oxides. Matovi´c et al. investigated the high-pressure behaviour of Multicomponent $(La_{1/8}Sm_{1/8}Nd_{1/8}Pr_{1/8}Y_{1/8}Gd_{1/8}Dy_{1/8}Yb_{1/8})_2(Hf_{1/2}Zr_{1/2})_2O_7$ system with pyrochlore structure experimentally and theoretically up to ~25 GPa[21]. They observed that the pyrochlore structure is stable up to ~25 GPa. The examined high-entropy pyrochlore showed varying peak intensities with pressure, which were ascribed to atomic interactions, crystal orientation, structural flaws, distances between planes, and crystal structure.

Recent research on conventional pyrochlores reveals that numerous compositions amorphized at high pressures, either immediately or after precursor crystal-to-crystal structural transition[18]. Several research groups have examined the high-pressure behavior of conventional pyrochlore and defect-fluorite $A_2B_2O_7$ zirconates, titanates, and hafnates[18,22]. Hafnates undergo a phase transformation from pyrochlore to cotunnite-type structure under high pressure between 18 and 25 GPa, and the transition is not completed up to 50 GPa. Upon the release of the pressure, hafnates changed into the defect-fluorite structure with partial amorphization[23]. However, conventional zirconates and titanates are known to undergo structural phase transformation around 16 and 42 GPa, respectively, as shown in **Table 1**[18,19,24,25]. The order-disorder occupancy of cations is the primary mechanism controlling the transformation of zirconate pyrochlores into a monoclinic or cotunnite-like phase under high pressure[26,27]. Transition pressure has been found to increase when actinide with large cationic radii is substituted on the A-site in pyrochlores, increasing the cationic radii ratio $(r_A/r_B)$[25]. The majority of past high-pressure investigations on conventional pyrochlores have mostly concentrated on the effect of changing cationic species to either the A or B-site or on both cationic sites[18,25,28].

Table 1. The behavior of Zirconate pyrochlores under high pressures. The $r_A/r_B$ values are from the Shannon radii database. "-" is not given in reference.

| A-site | Gd [27] | Nd [27] | Ce [27] | Gd [29] | Eu [18] | Dy [18] | Sm [30] | La [26] | La$_{0.5}$Gd$_{1.5}$ [26] |
|---|---|---|---|---|---|---|---|---|---|
| $r_A/r_B$ | 1.46 | 1.54 | 1.58 | 1.46 | 1.48 | 1.426 | 1.498 | 1.61 | 1.499 |
| $P_t$ (GPa) | 18(2) | 19(2) | 19(2) | 3 | 15.9 | 16.5 | 19.7 | 22.7 | 23.3 |
| HP-Phase | Monoclinic | Monoclinic | Monoclinic | Monoclinic | Cotunnite | Cotunnite | Distorted fluorite | Cotunnite | Cotunnite |
| $B_o$ (GPa) | 155(6) | 143(6) | 253(6) | 161.5(2.2) | - | - | - | 235(4) | 360(23) |
| $B_o'$ | 7(1) | 11(1) | 0.0 | 9(0.8) | - | - | - | 4(fixed) | 4(fixed) |

Apart from these studies, Srihari *et al.* studied the effect of particle size on the high-pressure behaviour of Yb$_2$Hf$_2$O$_7$ pyrochlore and found that a sample with a smaller particle size (~20 nm) exhibits superior structural stability and higher bulk modulus at high pressures as compared to the large particle size samples [31]. The increased incompressibility for smaller-sized nanoparticles could be due to several factors, including a high surface-to-bulk atom ratio that changed the average bond length, bond energy, and effective coordination environment[32]. In the present study, we systematically investigate the effect of particle size on the stability of high entropy (La$_{0.2}$Nd$_{0.2}$Sm$_{0.2}$Gd$_{0.2}$Yb$_{0.2}$)$_2$Zr$_2$O$_7$ ceramics at high pressures.

## 2. Materials and Methods

### 2.1 Synthesis of HEZ nanopowders

The polyacrylamide gel method was used for the first time to successfully synthesize the single-phase high entropy (La$_{0.2}$Nd$_{0.2}$Sm$_{0.2}$Gd$_{0.2}$Yb$_{0.2}$)$_2$Zr$_2$O$_7$ nanopowders. Rare-earth nitrate hexahydrates RE(NO$_3$)$_3$·6H$_2$O (RE=La, Nd, Sm, Gd, and Yb) with a high purity of 99.9%, zirconium oxychloride (ZrOCl$_2$·8H$_2$O, 99.9% purity), acrylamide (AM; C$_3$H$_5$NO, 99.0% purity), N-N'-methylene-bis-acrylamide (MBAM; C$_7$H$_{10}$N$_2$O$_2$, 99.0% purity), and ammonium persulfate ((NH$_4$)$_2$S$_2$O$_8$, 99.0% purity) were used as starting reagents for the synthesis of HEZ nanopowders. First, rare-earth nitrate hexahydrates and zirconium oxychloride were mixed with distilled water according to the stoichiometry of (La$_{0.2}$Nd$_{0.2}$Sm$_{0.2}$Gd$_{0.2}$Yb$_{0.2}$)$_2$Zr$_2$O$_7$ nanopowders. A clear solution was obtained after 30 minutes of vigorous stirring, and Zr$^{4+}$ concentration was 0.02 mol·L$^{-1}$. Subsequently, AM, MBAM, and (NH4)$_2$S$_2$O$_8$ in a molar ratio of 24/2/1 were introduced to the formed clear solution, thoroughly dissolved through vigorous stirring. In the mixed solution, the molar ratios of AM/$Zr^{4+}$ were changed to 20/1, 40/1, and 80/1. The combined solution was then heated to 80 °C in a water bath until it became a wet gel. After 3 days at 80 °C in the oven, the wet gel was dried to produce xerogel. Following xerogel calcination at 900–1300 °C for two hours in the air, which leads to the particle size of ~25(9) nm (900 °C) [HEZ-25nm], ~45(15) nm (1100 °C) [HEZ-45nm], and ~ 68(18) nm (1300 °C) [HEZ-68nm] were finally prepared. The average particle size of the samples was determined from TEM measurements. More details about the synthesis of HEZ nanocrystalline powders and microstructural analysis are reported elsewhere[33].

## 2.2 Experimental Techniques

Room temperature high-pressure powder XRD measurements were performed up to ~30 GPa using a Mao bell-type diamond anvil cell (DAC)[34], having a culet size of ~ 400 μm at the ECXRD Beamline (BL-11)[34] at Indus-2 synchrotron radiation source using monochromatic X-rays of wavelength 0.7305 Å. The sample chamber was made by drilling a ~ 200 μm diameter hole in a pre-indented tungsten gasket, indented from an initial thickness of ~ 250 μm to about ~50 μm. The fine powder samples were loaded with silicone oil as a pressure-transmitting medium and a few specks of gold powder, which served as a pressure indicator. The well-known equation of state for Au was used to calculate the pressure[35]. All the diffraction images were recorded in a MAR345 imaging plate detector. Diffraction images of standard reference materials, notably $CeO_2$ and $LaB_6$, were used to calculate the sample-to-detector distance, detector tilt, rotation, and calibration parameters. The acquired calibration parameters were used to convert the two-dimensional Debye ring pictures captured on the imaging plate into a one-dimensional diffraction profile using the FIT2D[36] and Dioptas[37] software programs. The GSAS-II[38] program was used to refine the 1D diffraction patterns according to the defect fluorite (DF) and pyrochlore (PY) structure. EOSFit7GUI software[39] was used to fit the pressure-volume (P-V) data to a second-order Birch-Murnaghan equation of state[40].

## 3. Results and Discussion

### 3.1 Crystal Structure at Ambient Pressure

The XRD patterns of nanocrystalline $(La_{0.2}{}^{Nd}{}_{0.2}Sm_{0.2}Gd_{0.2}Yb_{0.2})_2Zr_2O_7$ zirconates (HEZ) calcined at different temperatures depicted the distinct behavior in phase formation as shown in **Fig. 1 (a).** The sample sintered at 900 °C, having a lower particle size (HEZ-25 nm), exhibits the broad-peaked defect fluorite (DF) phase (Spacegroup Fm-3m), which lacks superstructure ordering. The peaks at 2θ values of 13.95°, 16.10°, 22.78°, 26.75°, 27.95°, and 32.36° can be well indexed as (111), (200), (220), (311), (222), and (400) reflections of the defect fluorite structure, respectively. The significant broadening of diffraction peaks can be attributed to smaller particle sizes and defects. The decreasing microstrain (~0.0282) at a lower sintering temperature (900 °C) compared to the microstrain (~0.0136) at a higher sintering temperature (1300 °C) suggests an improved crystallinity and reduction in the lattice defects. The diffraction peaks of the sample calcinated at a higher temperature (1100 °C) are sharper and shifted to lower angles, suggesting lattice parameter expansion and enhanced crystallinity with higher calcining temperatures. The HEZ samples sintered at 1100 °C and 1300 °C exhibit the strong diffraction peaks of defect fluorite (DF) and its superstructure pyrochlore (PY) phase. The phase fraction of the PY phase increases with sintering temperature, as is evident from the increased relative intensity of diffraction peaks corresponding to the PY phase. The structural parameters of the DF and PY phase, as obtained from the Rietveld refinement of diffraction patterns, are presented in **Table 2**. A crystal structure of DF and PY phases is shown in **Fig. 1(b)**. The DF structure lacks the ordered 2×2×2 superstructure found in the ordered PY, and the lattice parameter of DF is nearly half that of an ordered PY, as shown in **Table 2**. By randomly distributing both cations and anions onto their respective sublattices, the ordered pyrochlore structure can be changed to the disordered defect-fluorite structure, with a rise in A-site coordination number (CN) from 8 and 6 to an average of 7 for both [41]. The A- and B-site cations occupy the 16d (1/2, ½, ½) and 16c (0, 0, 0) Wyckoff sites. The 48f ($x_{48f}$, 1/8, 1/8) site has the O(1) oxygen ions and the 8b (3/8, 3/8, 3/8) site contains the O(2), while the 8a (1/8, 1/8, 1/8) sites are vacant. The DF structure is a face-centered cubic array of cations at the 4a Wyckoff position, where oxygen atoms are positioned at every tetrahedral interstice that is accessible; at the 8c Wyckoff position (1/8th of the oxygen locations are randomly unoccupied) [42]. Five distinct rare earth elements on the A-site with varied radii were selected for the presented study. The higher radii of A-site cations ($r_A$) stabilize the PY structure, which enables sufficient filling of the vast voids in the lattice. Nonetheless, the B-site elements are usually smaller than the A-site cations, and their reduced radii ($r_B$) enable the PY structure to establish strong bonds with oxygen. The cationic radii ratio ($r_A/r_B$) significantly influences the crystal structure of pyrochlores. In the present study, the $r_A/r_B$ value of 1.52 indicates the formation of a stable PY structure at high temperatures, which generally forms when $r_A/r_B$ falls within the range of 1.46–1.78. Although it is an essential thermodynamic criterion for pyrochlore stability, the $r_A/r_B$ ratio is insufficient on its own to ensure that it will form at lower temperatures. As a result, the DF phase is stabilized at lower sintering temperatures, and the PY phase is generated at higher sintering temperatures.

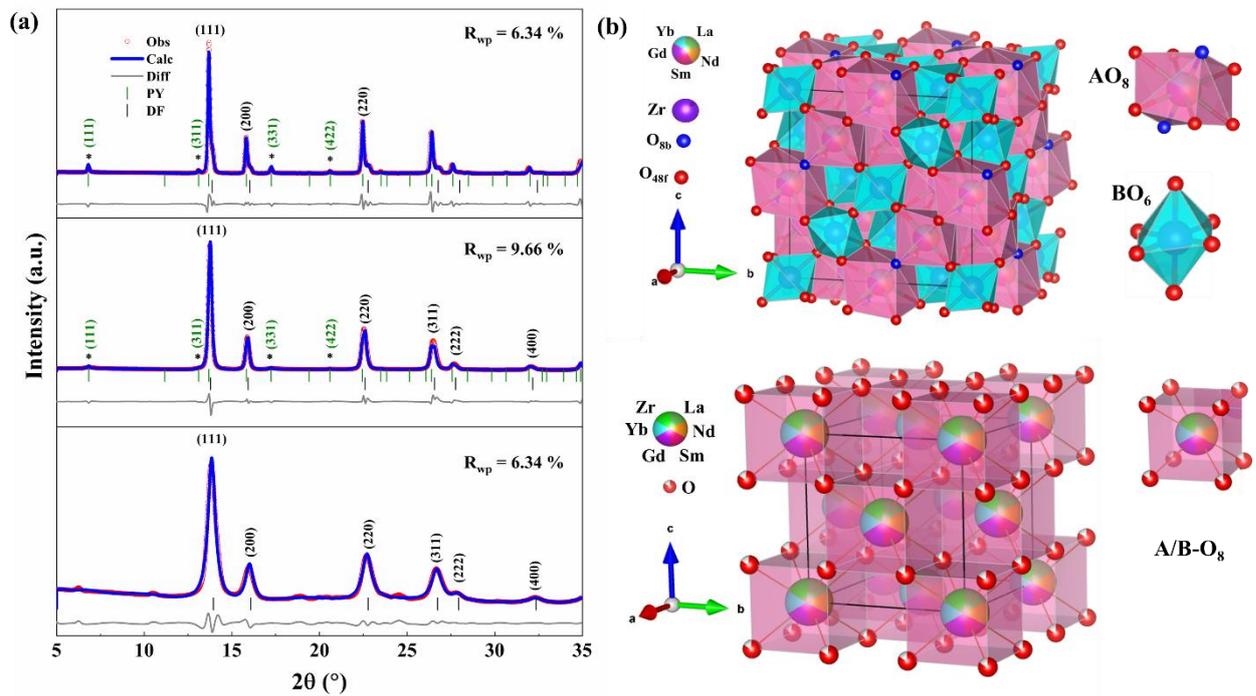

**Fig. 1.** (a) Refined XRD patterns of as-synthesized high entropy $(La_{0.2}Nd_{0.2}Sm_{0.2}Gd_{0.2}Yb_{0.2})_2Zr_2O_7$ nanopowders having different particle sizes, namely, HEZ-25nm, HEZ-45nm, and HEZ-68nm, (b) Visualization of the crystal structure for ordered pyrochlore ($A_2B_2O_7$, Fd-3m) and disordered defect fluorite ($(A, B)_4O_7$, Fm-3m).

Long-range ordering of cations and oxygen vacancy processes is necessary for pyrochlore production, whereas atomic diffusion is constrained in the DF structure at lower sintering temperatures. Especially interesting is the situation in which a high-entropy pyrochlore structure is formed when many elements occupy the same site [21].

**Table 2**: The structural parameters obtained from the refinement of ambient XRD patterns for $(La_{0.2}Nd_{0.2}Sm_{0.2}Gd_{0.2}Yb_{0.2})_2Zr_2O_7$ nanopowders with different particle sizes.

| Sample | Lattice Parameter (Å) | | Theoretical density (DF) | $x_{O48f}$ (PY) | Residuals $R_{wp}$ (GSAS) |
|---|---|---|---|---|---|
| | DF | PY | | | |
| **HEZ-25 nm** | 5.2720 (2) | - | 7.88 | - | 6.34 % |
| **HEZ-45 nm** | 5.2651 (3) | 10.5955 (6) | 7.91 | 0.3238 (17) | 9.66 % |
| **HEZ-68 nm** | 5.2426 (6) | 10.6193 (2) | 8.01 | 0.3270 (14) | 6.34 % |

All the above results indicate that high entropy $(La_{0.2}Nd_{0.2}Sm_{0.2}Gd_{0.2}Yb_{0.2})_2Zr_2O_7$ nanopowders exhibiting DF and PY structures have been successfully synthesized by calcining the HEZ nanopowders at different temperatures.

**3.2 In-situ High-Pressure Synchrotron ADXRD Study**

Representative XRD patterns of HEZ-25nm, HEZ-45nm, and HEZ-68nm are shown in **Fig. 2** up to ~ 30 GPa. The diffraction peaks of all three samples broaden, weaken, and gradually shift to a higher angle with increasing pressure, indicating compressive volumetric strain accompanied with structural distortion. The XRD pattern of HEZ-25nm shows that all the DF diffraction peaks, such as (111), (200), and (220), are still present up to the highest pressure of ~30 GPa, suggesting stability of the DF phase up to highest measured pressure. A broad hump beneath the strongest diffraction peak (111) corresponding to diffuse scattering from the distorted/amorphous phase emerges above ~24 GPa at grows with subsequent higher pressures. The broadening of the strongest diffraction peak (111) of the DF phase starts above ~ 23 GPa and ~15 GPa in the HEZ-45 nm and HEZ-68 nm samples, respectively. The intensity of the pyrochlore superstructure peak (111) and (311) gradually disappear with increasing pressure, indicating the loss of the PY phase with pressure. Although very weak, the PY diffraction peaks are discernible up to the highest measured pressure along with DF diffraction peaks, and no new peaks emerge in HEZ-45 and HEZ-68nm samples. These variations can be ascribed to atomic interactions, atomic disordering, crystal orientation, structural defects, distances between surfaces, and crystal structure [1].

Earlier reports show that conventional zirconates typically undergo orthorhombic cotunnite-like structure or monoclinic phase transformations under high pressure, as shown in **Table 1**. Tanate pyrochlores usually undergo a monoclinic phase transformation, whereas zirconate pyrochlores usually undergo an orthorhombic phase transformation[18,28,43]. The typical cotunnite structural model cannot adequately fit the high-pressure phase of pyrochlores. The order-disorder transition in oxides with pyrochlore structure highly depends on the cation size ratio on the A- and B-sites. A fully ordered pyrochlore structure is stable when the size ratio $(r_A/r_B)$ is greater, as in rare earth titanate sand stannates [44]. The smaller size ratio ($r_A/r_B$ <1.46) is conducive to the occurrence of cation anti-site defects, and a disordered fluorite structure is stable for zirconates at ambient conditions.

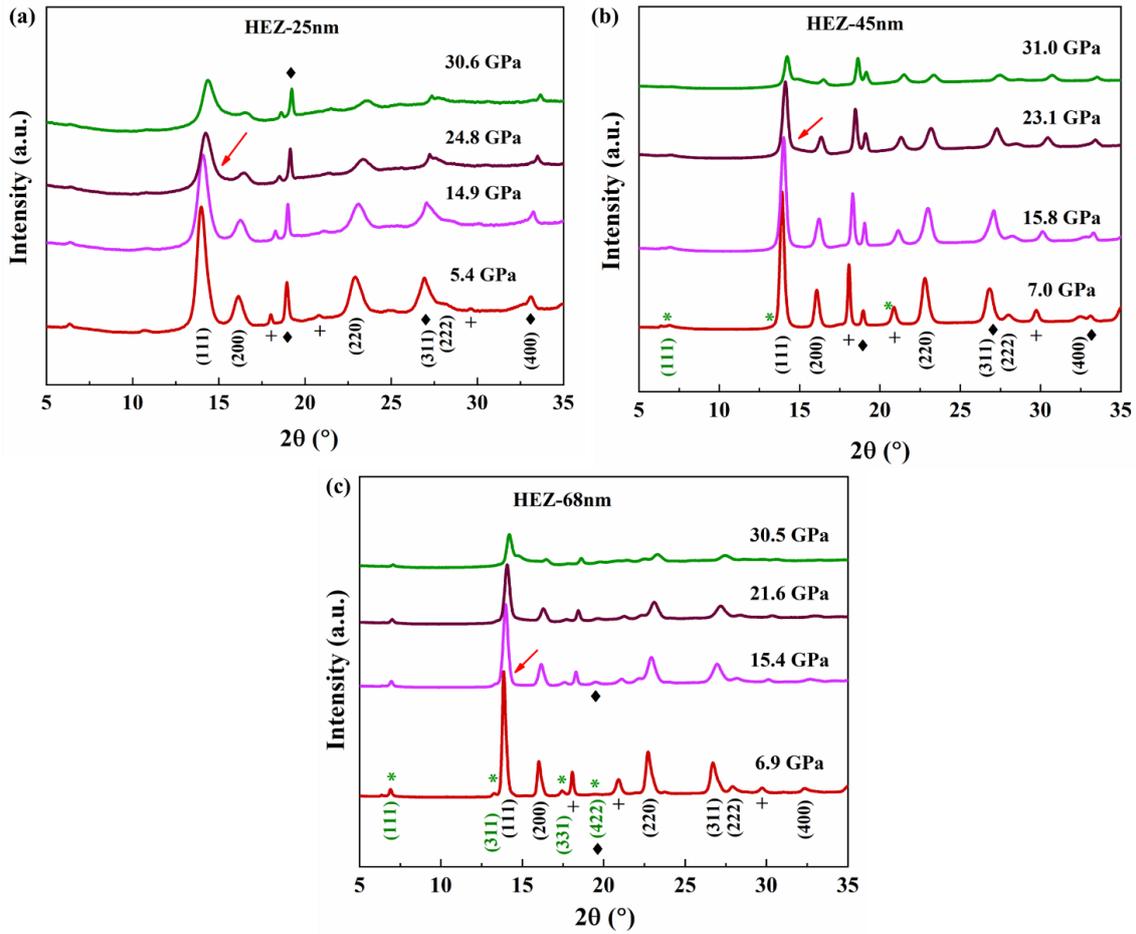

**Fig. 2** High-pressure ADXRD patterns of HEZ-25nm, HEZ-45nm, and HEZ-68nm nanopowders recorded at various pressures using silicone oil as the pressure-transmitting medium. The arrow mark shows the broadening of the most intense (111) peak of the DF phase at high pressure. The diffraction peaks marked with '*' (Green color) show the appearance of PY superstructure peaks, '+' symbols correspond to the pressure marker, gold, and '♦' correspond to the gasket material, tungsten, respectively.

In contrast, the investigated high-entropy zirconates in this study do not undergo any structural phase transition up to ~30 GPa and show the stability of their initial DF and PY structures. The PY phase fraction for the HEZ-45 and HEZ-68 nm samples decreases as pressure increases. Since there are no new peaks and the intensity decreases with pressure, it is likely that these materials are approaching the DF phase or amorphization and losing their pyrochlore structure as the pressure increases. The current investigations are in good agreement with the earlier high-pressure work on high entropy $(La_{1/8}Sm_{1/8}Nd_{1/8}Pr_{1/8}Y_{1/8}Gd_{1/8}Dy_{1/8}Yb_{1/8})_2(Hf_{1/2}Zr_{1/2})_2O_7$ pyrochlore which demonstrated the stability of the PY structure up to ~25 GPa [21]. Our earlier investigation on titanate pyrochlore, which was linked to the disordering of the polyhedrons, showed a similar type of intensity decrement with increasing pressure [42]. The structural stability of high entropy oxides (HEOs) is promoted by the tendency of the high entropy resulting from multiple elements at each specific site to reduce the free energy. Conversely, lattice distortion tends to increase the free energy by interrupting the ideal crystal lattice. This interaction between lattice distortion and entropy-driven stability highlights how HEO structures can be modified [45]. Liu *et al.* investigated the mechanical properties of the DF $(La_{0.2}Nd_{0.2}Sm_{0.2}Gd_{0.2}Yb_{0.2})_2Zr_2O_7$ (Particle size ~210 nm) and observed the enhanced mechanical properties, including higher fracture toughness and large Young's modulus compared to the conventional zirconates with a single element at each site [46]. The multi-phase refinement of the XRD pattern of all three samples was performed up to ~ 30 GPa to confirm the high-pressure phase, as demonstrated in **Fig. 3**.

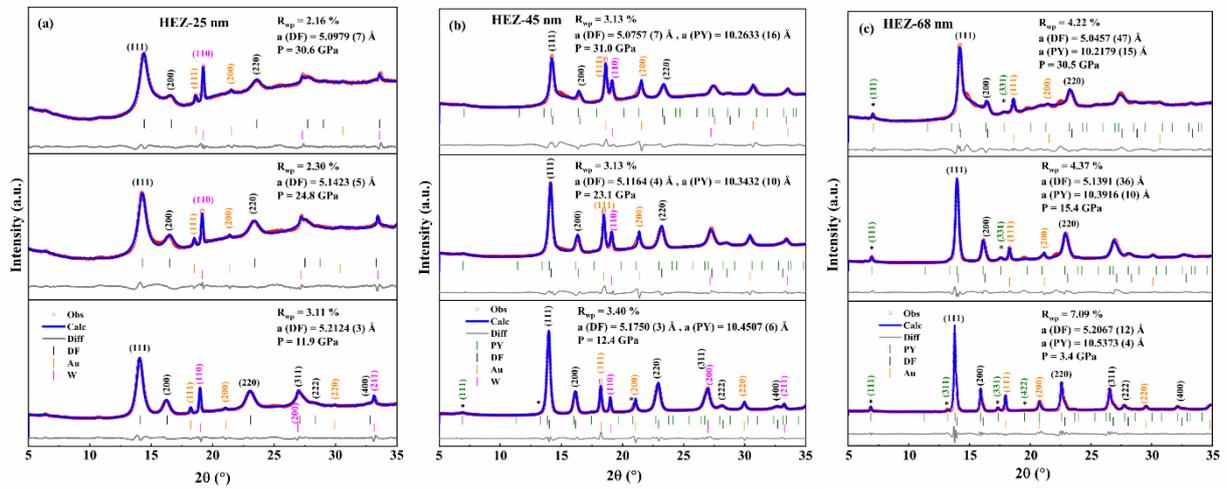

**Fig. 3.** Refined XRD patterns of samples (a) HEZ-25 nm, (b) HEZ-45 nm, and (c) HEZ-68 nm at different pressures. Observed (hollow red circles) and calculated (solid blue lines) diffraction patterns at different pressures. Plotted at the bottom is the difference curve. Tick marks with black, olive, orange, and magenta are the estimated reflection positions of the DF, PY, Gold (Au), and Tungsten (W) phases.

We refined the XRD patterns using both DF (Fm-3m) and PY (Fd-3m) up to the highest pressure to confirm the mixing of these phases, and the refined results are presented in **Fig. 3**. This process involved refining the lattice parameters, atomic locations, lineshape, and background. As shown in **Fig.3**, the agreement factors ($R_{wp}$) show that the quality of the refinements remained high throughout the three samples. **Fig. 4** plots the unit cell volume of the DF and PY phases as a function of pressure utilizing the Rietveld refinement of the XRD patterns recorded from ambient to ~30 GPa. The P-V plot of both the phases in all the samples exhibits anomalous plateau region between 7 to 15 GPa. The P–V curve was fitted to the second-order Birch-Murnaghan equation of state (EoS) [40] in two regions, viz. ambient to 7 GPa and 15 to the highest measured pressure, avoiding the anomalous region, to obtain the bulk modulus of all three samples.

$$P = \frac{3.0}{2} B_o \left[ \left(\frac{V_o}{V}\right)^{\frac{7}{3}} - \left(\frac{V_o}{V}\right)^{\frac{5}{3}} \right] \times \left( 1 + \frac{3.0}{4}(B'_o - 4) \times \left[ \left(\frac{V_o}{V}\right)^{\frac{2}{3}} - 1 \right] \right) \quad (1)$$

where $B_o$ is the bulk modulus, $B'_o$ is the derivative of the bulk modulus, which is set to 4 for the second-order equation, P is the pressure, and $V_o$ is the unit-cell volume at room temperature and ambient pressure.

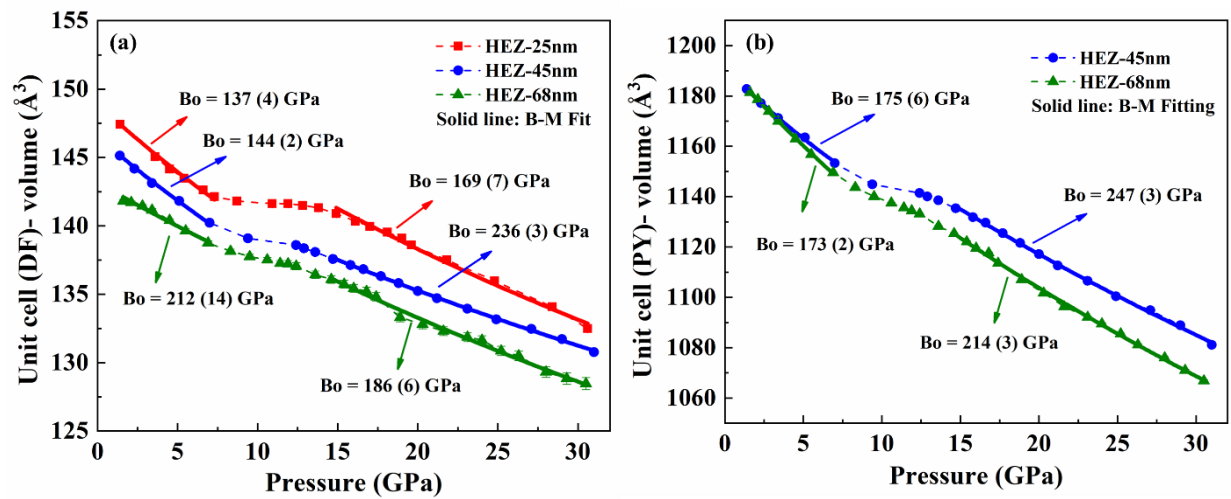

**Fig. 4.** The pressure dependence of the unit cell volume of (a) DF phase and (b) PY phase. The solid lines correspond to the B-M EoS fitting for all three samples in different pressure regions.

Our results indicate the dispersed behavior of the P-V curve in the different pressure regimes for all three samples. Two data regions were used for the fits for both DF and PY phases: one up to 7 GPa to reduce the impact of non-hydrostatic influences on the results above 15 GPa. The bulk modulus for the HEZ-25 nm sample shows a systematic increase in the particle size for the DF phase up to ~7 GPa. The obtained bulk modulus of DF phases is ~137(4) GPa, 144(2) GPa, and 212 (14) GPa for the HEZ-25nm, HEZ-45nm, and HEZ-68nm, respectively. This indicates that larger particle size samples (~68 nm) are more uncompressible, up to ~7 GPa, compared to the lower particle size sample (~25nm). The bulk modulus for the DF phase above 15 GPa is ~169(7) GPa, ~236(3) GPa, and ~186(6) GPa for the HEZ-25nm, HEZ-45nm, and HEZ-68nm, respectively. The bulk modulus of PY phases up to ~7 GPa are ~175(6) GPa and ~173(2), respectively. The bulk modulus for the PY phase above 15 GPa is ~247 (3) GPa and ~214 (3) GPa. Our bulk modulus value of the DF phase for larger particle size zirconate sample [~68(18) nm] is 212 (14) GPa up to ~7 GPa, which agrees with the bulk modulus value of ~210 (10) GPa for DF structured hafnate composition of particle size ~72(22) nm[31]. The bulk modulus of the PY phase of HEZ-45nm and HEZ-68 nm samples also show higher values than conventional pyrochlores such as $Nd_2Zr_2O_7$ $Gd_2Zr_2O_7$ compositions[27]. In our previous study, [42] on $GdYTi_2O_7$ pyrochlore, a systematic variation of unit cell volume with increasing pressure was observed, while an unusual behavior of unit cell volume is seen in the present study of multicomponent HEZ. The unusual behavior is more evident for the lower particle-size sample (HEZ-25 nm) than the higher particle-sized HEZ-45 and HEZ-68 nm samples. A similar anomalous P-V behavior has been earlier reported in conventional pyrochlores such as $La_2Zr_2O_7$ [19], $Sm_2Zr_2O_7$ [30,] and $Eu_2Sn_2O_7$ [44]. When methanol/ethanol with or without water was employed as a pressure-transmitting medium (PTM), a slope in the P-V curve of $La_2Zr_2O_7$ was seen even with the anomalous lattice expansion at ~10 GPa; this was not the case when Ar was used as PTM. The study implied that the water molecule from the ethanol (PTM) interacts with the PY structure, resulting in the slope observed in the P-V curve. Similar abnormalities in the P-V curve of $Eu_2Sn_2O_7$ above ~23 GPa also point to the intercalation of water (from ethanol) into the pyrochlore structure. The incorporation of water or hydroxyl in the pyrochlore structure at high pressures may be common in pyrochlore oxides. Perottoni *et al.* studied the pressure-induced water insertion in the defect PY $NH_4NbWO_6$ using different PTM, including NaCl, methanol-ethanol (4:1), methanol-ethanol-water (16:3:1), and silicone oil [47]. The sudden increase in the lattice parameter with unusual behavior of the P-V curve and pronounced abnormalities in the Raman shift were observed when the PTM was methanol-ethanol-water (16:3:1). Silicone oil does not produce similar abnormalities in the same composition when used as the pressure medium. According to their initial research, which was based on electrostatic calculations, the oxygen of the water molecule most likely occupies the unoccupied 8b site in the PY crystal [47]. In some cases, even silicone oil as PTM also results in the same tendency in the P-V curve; for example, $Gd_{1.5}Ce_{0.5}Ti_2O_7$ PY composition when loaded with silicone oil, anomalous P-V behaviour is observed in the pressure range ~6.5-13 GPa [25]. The observed changes were associated with the non-hydrostatic conditions arising during the solidification of PTMs. A similar plateau in the P-V curve was observed above ~5 GPa in a recent high-pressure investigation on $Lu_2Ti_2O_7$ and $Lu_{1.5}Ce_{0.5}Ti_2O_{7+x}$ pyrochlores using silicone oil as PTM [48]. The anomalous P-V curves are frequently documented in pyrochlores, but the root cause of these unusual changes is still unclear. Whether it is pressure-induced intercalation or solidification, the PTMs certainly have some impact on the high-pressure behaviors of pyrochlores. Another possible explanation for this unusual behavior may be the compressive stresses in the core and tensile stresses at the surface, which may play a competitive role. As the surface-to-surface-to-volume ratio is the largest for the HEZ-25 nm sample, the anomalous effect is more prominent in this case. The PV plots in the present study show a systematically decreasing volume of the DF structure with increasing particle size. For the PY phase in HEZ-45nm and HEZ-68nm samples, the initial unit cell volume is similar, but the HEZ-68nm sample is relatively more compressible than the HEZ-45nm sample. This suggests that tensile stresses predominate in samples of smaller particle sizes.

The Rietveld refinements also obtained the pressure dependence of the bond length and volumes of the polyhedrons connected to the DF and PY structures. **Fig. 5** shows the variation in the bond length and polyhedrons of the DF and PY phases calculated from refining the XRD patterns with increasing pressure.

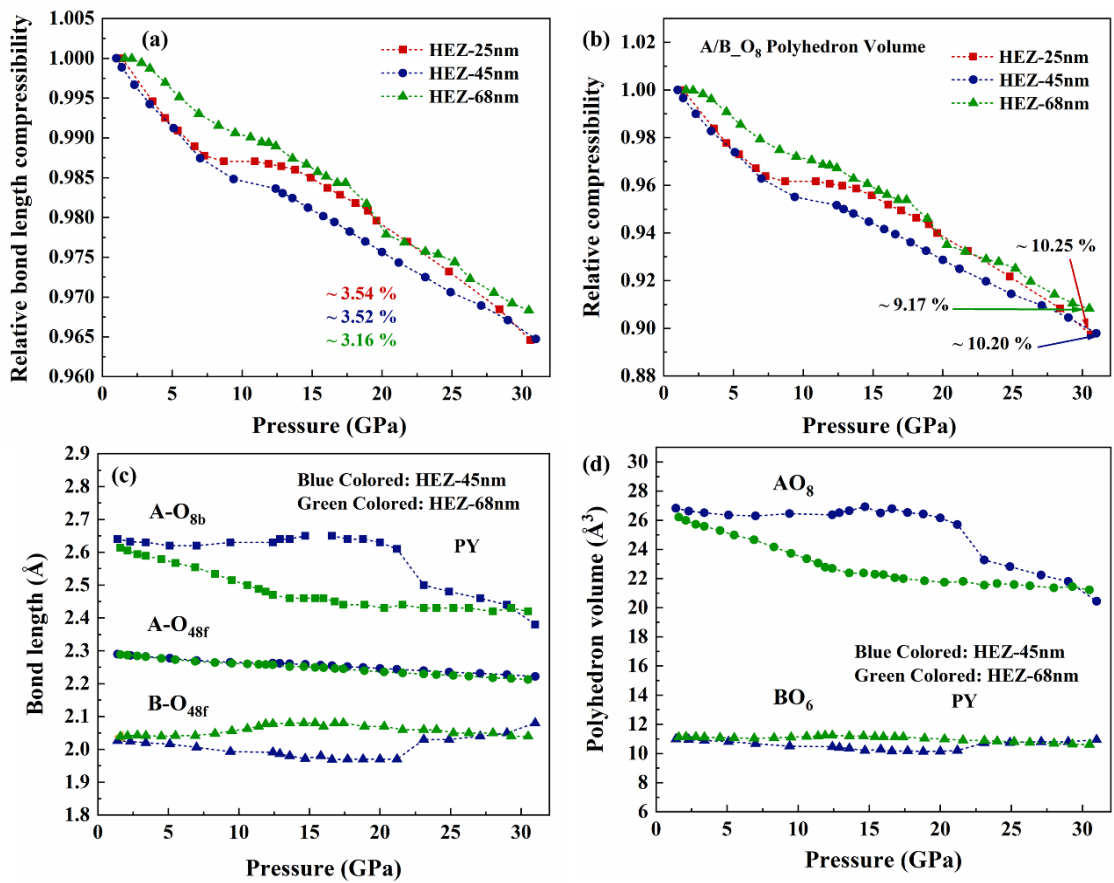

**Fig. 5.** Calculated relative (a) cation-anion bond lengths, (b) volume compressibility of $A/BO_8$ polyhedron of the DF phase, (c) Cation-anion bond length of the PY phase and (d) polyhedron's volume of both $AO_8$ and $BO_6$ polyhedrons of PY phase of the high entropy $(La_{0.2}Nd_{0.2}Sm_{0.2}Gd_{0.2}Yb_{0.2})_2Zr_2O_7$ nanopowders system versus pressure.

The DF structure only contains one polyhedron (A/B-$O_8$) since the A and B-site cations share a crystallographic location. The A/B-O bond length and polyhedron volume show fewer changes for the HEZ-68 nm sample than the HEZ-25nm and HEZ-45nm. It is clear that A–O bonds are more compressible than B–O bonds for the PY-structured HEZ-45 nm and HEZ-68 nm samples; as a result, the compression of $AO_8$ dodecahedra is primarily responsible for the notable volume reduction seen under pressure. The compressibility of the unit cell lies between the compressibilities of these two polyhedrons [42]. Certain bond lengths, such as A-$O_{8b}$ and B-$O_{48f}$, also exhibit anomalous behavior, as seen in the P-V curve of all three samples. This behavior is particularly pronounced for the HEZ-45 nm sample. Since the < Zr–O > bonds are less compressible than the < A–O > bonds, the < A–O > bonds are more likely to be modified when compressed. These findings imply that the < Zr–O > bonds ought to be more significant than the < A–O > bonds in the stability of the PY structure at high pressure. The robustness of the B-O in the high-pressure stability of PY is also reported on some conventional pyrochlores [42,49]. The degree of anion disordering can be quantitatively scaled using $x_{48f}$, the only changeable atomic coordinate in the pyrochlore structure. **Fig. 6**. Shows the variation phase fraction of PY structure and change in the positional coordinate $x_{48f}$ with increasing pressure.

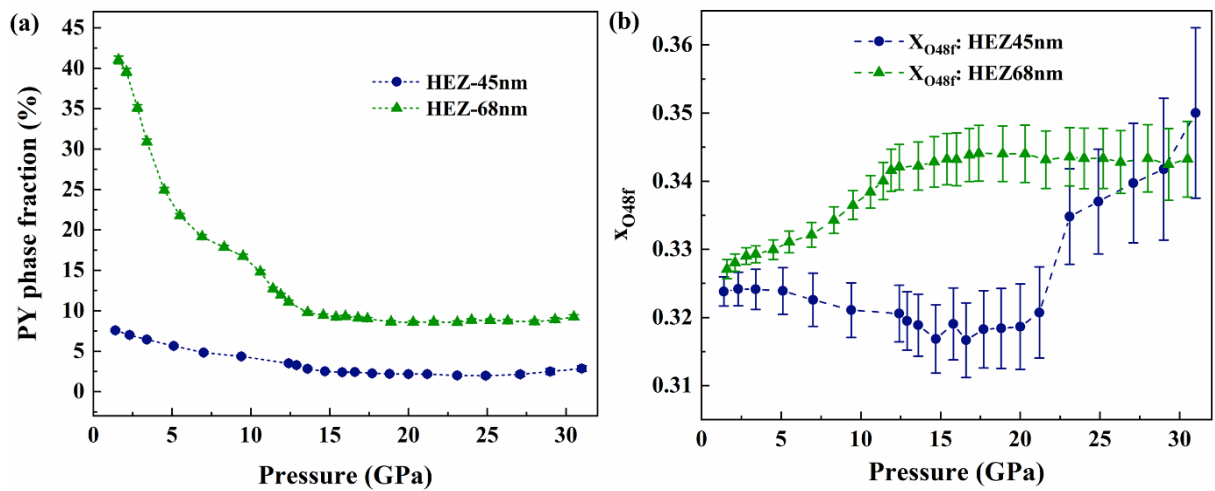

**Fig. 6.** Calculated (a) Phase fraction of PY and (b) Variation in the x-parameter of $O_{48f}$ anions ($x_{O48f}$) with the pressure for the PY-structured HEZ-45 nm and HEZ-68nm samples.

The HEZ-68 nm sample loses its original PY structure more rapidly than the HEZ-45 nm sample when the pressure increases. The XRD pattern of HEZ-68 nm shows more crystalline peaks of the PY superstructure lattice than HEZ-45 nm, which shows less crystallinity of the PY structure. PY phase decreases systematically with increasing pressure for HEZ-68 nm, whereas it changes less for HEZ-45 nm, as shown in **Fig. 6 (a)**. The PY decreases rapidly up to ~12 GPa, then decreases slowly with increasing pressure. Generally, the $x_{O48f}$ positional coordinate in pyrochlore shifts towards 0.3125 to maintain the cationic ordering, and in an ideal fluorite structure, the $x_{O48f}$ is 0.375. The value of $x_{O48f}$ increases more quickly up to ~12 GPa, then almost constantly up to a high pressure of ~30 GPa for HEZ-68 nm, as illustrated in **Fig. 6 (b).** For HEZ-45 nm, it increases more quickly above ~20 GPa. The value of $x_{O48f}$ increases from ~0.3238 to ~0.3500 and ~0.3271 to ~3432, up to the highest pressure of ~30 GPa for the HEZ-45nm and HEZ-68nm, respectively. This variation indicates that these samples are losing the PY structure and approaches to stabilizing the DF structure with increasing pressure. These results suggest the disordering of cations with increasing pressure and are consistent with observed disordering in $Eu_2Z_2O_7$ and $GdYTi_2O_7$ [42,49]. This disordering leads to the distortion of the polyhedrons, resulting in the decrement of the intensity of XRD patterns. Additionally, the isotropic thermal parameters of both cationic sites (16d and 16c) were independently refined using Rietveld analysis of the XRD patterns to support the disordering of the cations and distortion of the polyhedrons, as illustrated in **Fig. 7**.

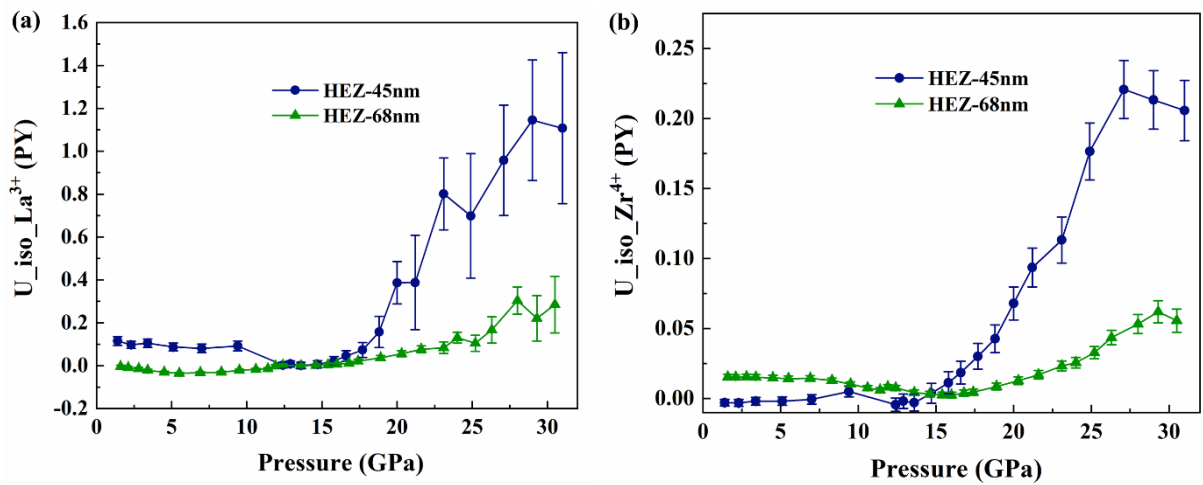

Fig. 7. Variation of refined isotropic thermal parameters with the hydrostatic pressure for both cations at (a) A-site and (b) B-site for PY structure.

As pressure increases, the refinement results demonstrate a relative rise in the thermal parameters for the A- and B-site cations, especially for the HEZ-45nm sample. The high degree of average positional disorder in the A and B-site cationic sublattice explains this increase in the disordering of cations with increasing pressure. Subtle anion/cation positional disordering and polyhedron distortions regarding their mean periodic lattice positions in the pyrochlore structure are the causes of the variations in peak intensities with increasing pressure. As a result, the preliminary investigation suggests that grain size is crucial for adjusting the phase stability of HEO materials under high pressure. Earlier research on nanosized materials indicates that a reduction in grain size causes a notable rise in bulk modulus compared to the bulk composition. This has been clarified by the increased contribution of surface energy in nanomaterials [31,50,51]. The improvement in physicochemical characteristics can be explained by the higher surface energy in samples with smaller particle sizes than in bulk ceramic samples. Compared to the larger particle sizes of the HEZ-45 and HEZ-68 nm samples, the smaller HEZ-25 nm sample in this study exhibits dominating anomalous P-V behaviour. The HEZ materials show superior phase stability and incompressibility than conventional zirconate materials under high pressure. The elevated compositional complexity and severe lattice distortion are typically the causes of the higher stability of high-entropy materials, similar to high-entropy alloys. The constituent multiple elements found in the matrix and several regulating parameters, such as the ionic radii and size of the elements, significantly impact the properties of these ceramics.

**Conclusion**

The angle-dispersive synchrotron X-ray diffraction observations were used to examine the high-pressure behavior of high entropy $(La_{0.2}Nd_{0.2}Sm_{0.2}Gd_{0.2}Yb_{0.2})_2Zr_2O_7$ nanopowders. We have found that the particle size significantly impacts the response to pressure, which agrees with earlier research. Ambient ADXRD results demonstrate the nanopowders sintered at 900 °C form in defect fluorite phase, which transformed to the pyrochlore phase with increasing sintering temperature to 1100 °C and 1300 °C. Ambient ADXRD results demonstrate that nanopowders sintered at 900 °C (HEZ-25nm) form in the defect fluorite (DF), then transition to the pyrochlore (PY) phase as the sintering temperature is raised to 1100 °C (HEZ-45nm) and 1300 °C (HEZ-68nm). The Rietveld refined XRD patterns at high pressure confirm the stability of ambient structures at high pressure. The experimental P-V data shows unusual behavior with the deviation from the EoS for HEZ, which is more evident for the lower particle-size sample (HEZ-25 nm) than the higher particle-sized HEZ-45 and HEZ-68 nm samples. The observed anomalous behaviour might be associated with the non-hydrostatic conditions arising due to the solidification of PTM or the interplay of compressive and tensile stresses in the core-shell regions of nanoparticles. This change in the $O_{x48f}$ positional coordinate shows that larger particle samples are losing the PY structure and are trying to stabilize the DF structure under rising pressure. For the PY-structured HEZ-45 nm and HEZ-68 nm samples, it is found that A–O bonds are more compressible than B–O bonds; hence, the volume loss observed under pressure is mainly caused by the compression of $AO_8$ dodecahedra. The refining results indicate that the changes in peak intensities with increasing pressure are due to polyhedron distortions concerning their mean periodic lattice positions in the pyrochlore structure and cationic disordering with increasing pressure. In contrast to conventional zirconates, HEZs do not undergo structural phase transition up to ~ 30 GPa and show the stability of DF and PY phases up to the highest measured pressure.

## CRediT authorship contribution statement

**Yogendar Singh:** Investigation (equal), Data curation (equal), conceptualization (equal), writing-original draft, Writing-review & editing, validation (equal), **Vivek Kumar:** Investigation (equal), Data curation (equal), conceptualization (equal)**, Xinghua Su:** Investigation (equal), Writing-review & editing**, Himanshu Kumar Poswal:** Data curation (equal)**,** Writing-review & editing**,** Investigation (equal), **K. K. Pandey:** Investigation (equal), Data curation (equal), conceptualization (equal), Writing-review & editing, **Pawan Kumar Kulriya**: Supervision, Funding acquisition, Validation (equal), Writing-review & editing (equal).

## Acknowledgments

The authors are thankful to the Raja Ramanna Center for Advanced Technology (RRCAT), Indore, for providing user support at ECXRD beamline-11, Indus-2 for high-pressure XRD measurements. Yogendar Singh acknowledges the Ministry of Education, Govt. of India, for awarding the Prime Minister's Research Fellowship (PMRF) with reference number PMRF-2122-2836. One of the authors, Pawan Kumar Kulriya, acknowledges the Board of Research in Nuclear Sciences (BRNS) for providing financial support under project 58/14/05/2019-BRNS/37013.

## Declaration of Competing Interest

The authors declare that they have no known competing financial interests.

## Data Availability

The data that support the findings of this study is accessible upon reasonable request from the corresponding authors.